\documentclass[review, compress]{elsarticle}
\usepackage{natbib}
\usepackage[ruled,vlined]{algorithm2e}
\usepackage{rotating}
\usepackage{hyperref}
\usepackage{amsfonts}
\usepackage{amsopn}

\usepackage{amsmath,cool}
\usepackage{hyperref}
\usepackage{amsmath}

\usepackage[english]{babel}
\usepackage{graphicx}
\usepackage{epstopdf}
\usepackage{geometry} 
\usepackage{fleqn}
\usepackage[rightcaption]{sidecap}

\usepackage{algorithmic}

\ifpdf
  \DeclareGraphicsExtensions{.eps,.pdf,.png,.jpg}
\else
  \DeclareGraphicsExtensions{.eps}
\fi
\usepackage{mathtools}

\usepackage{tikz}
\usetikzlibrary{quantikz}

\def\BibTeX{{\rm B\kern-.05em{\sc i\kern-.025em b}\kern-.08em
    T\kern-.1667em\lower.7ex\hbox{E}\kern-.125emX}}

\journal{Journal of \LaTeX\ Templates}

\begin{document}

\begin{frontmatter}

\title{A Hybrid Classical-Quantum framework for solving Free Boundary Value Problems and Applications in Modeling Electric Contact Phenomena
\tnoteref{mytitlenote}}


%\author{Merey M.Sarsengeldin\fnref{myfootnote}, Zuhair M. Nashed %\fnref{myfootnote}}
%\address{Department of Mathematics, University of Central Florida, Orlando, FL}

\author[mymainaddress,mysecondaryaddress]{Merey M.Sarsengeldin\corref{mycorrespondingauthor}}
\cortext[mycorrespondingauthor]{Corresponding author}
\ead{dr.sarsengeldin@gmail.com}


\address[mymainaddress]{Department of Mathematics, University of Central Florida, Orlando, Fl, US}
\address[mysecondaryaddress]{National Academy of Sciences, Institute of Mathematics and Mathematical Modeling, Almaty, Kazakhstan}
\begin{abstract}
In this paper we elaborate a hybrid classical-quantum framework which allows one to model and solve heat and mass transfer problems occurring in electric contacts. We utilize special functions and Harrow-Hassidim-Lloyd (HHL) quantum algorithm for finding temperature and arc flux functions exactly and approximately for the Stefan type problems. The Stefan type problems we are considering  
are based on Generalized Heat Equation with free boundaries. As examples we consider exact and approximate solutions of inverse one-phase and two-phase Stefan problems. An Inverse Generalized One-Phase Stefan Problem is considered as a model problem. Computational experiments were conducted and demonstrated on IBM Quantum Machine.  %and we demonstrate adequacy and agreement of the developed model with experimental data, which was developed on the base of elaborated mathematical framework. 
\end{abstract}

\begin{keyword}
Stefan Problem, Free Boundary Value Problems, Quantum Computing, HHL quantum algorithm, Electric Contacts, Arc Heating.
\end{keyword}

\end{frontmatter}

%\linenumbers

\section{Introduction}

%Mathematical modeling of Heat and Mass transfer problems in diverse phenomena  play crucial role []. Particularly in electric contact phenomena [], processes like Arcing, Bridging are so fast that mathematical modeling can be the only approach to predict and circumvent unwanted consequences []. Therefore, development of mathematical models and new computational frameworks is still important.\\
%From theoretical point of view, Heat and Mass transfer problems with phase transformations can be modeled by Stefan type problems where along with solution of system of partial differential equations moving boundary(s) has to be determined []. However, in cases with degenerate domains where moving boundary degenerates at the initial point additional difficulties occur due to the singularities in equations. Therefore, developing mathematical methods for solving  such kind differential equations is very topical.\\

In this study we develop a hybrid classical-quantum framework which allows one model heat and mass transfer problems occurring in electric contact phenomena. Along with the agreement with the experimental data this framework allows one efficiently find temperature function, arc flux, and unknown moving phase transformation boundary functions using quantum algorithms.
In this particular study we utilize Harrow Hassidim Lloyd (HHL) \cite{hhl09} algorithm for finding coefficients of heat temperature functions represented in the form of heat polynomials \cite{Haimo,Haimo1,Widder, nasim_aggarwala_1984} and arc flux function.\\

Pioneering investigations \cite{ben80,fe82} in 1980s brought a new computing paradigm known as quantum computing, in which information is encoded in a quantum system. In the 1990s, a series of works \cite{sho97,gro96,gro97} dedicated to quantum algorithms showed that they could do tasks faster than the best-known classical algorithms. Consistent advances in theory and experiments have resulted in a plethora of powerful quantum algorithms \cite{mo16} that outperform their classical counterparts in terms of computational power; however, due to the challenges associated with their physical realization, their applications are limited to a few use cases. Careful physical realization of algorithms on near term quantum computers \cite{Preskill2018quantumcomputingin, Ritter_2019} might lead to profound results in computational speed-up \cite{Supremacy, Harrow_2017}.

%We will be using in our study one of such powerful quantum algorithms developed by Harrow-Hassidim-Lloyd (HHL) \cite{hhl09} to solve MBVPs. The HHL algorithm consists of three steps: phase estimation, controlled rotation and uncomputation and is the special case of the continuous variable operator inversion quantum algorithm which can be used for solving PDEs, ODEs and widely used in quantum machine learning. A system of non homogeneous differential equations or system of linear algebraic equations can be represented in the following form

In this paper, we will study Free Boundary Value Problems (FBVPs) using one of the powerful quantum algorithms proposed by Harrow-Hassidim-Lloyd \cite{hhl09}. The HHL algorithm is widely used in quantum machine learning, for solving PDEs, ODEs \citep{Montanaro_2016,Wiebe,Lloyd, Rebentrost} and was physically realized on different quantum architectures \citep{Barz2014-fo, Cai, Zheng, Pan_PhysRevA}. It consists of three steps: phase estimation, controlled rotation, and uncomputation. The FBVPs are reduced to following equation where we apply HHL algorithm
\begin{align}
M\ket{x} = \ket{b},
\label{mateq}
\end{align}
In this study we assume that $M$ is Hermitian and sparse matrix, and b is a vector column. This condition can be relaxed and it can be shown that $\tilde{M} = \begin{bmatrix}
0 & M\\
M^{T} & 0
\end{bmatrix}$ can be brought to Hermitian matrix.
Since $\tilde{M}$ is Hermitian, we can solve the equation 
$\tilde{M}\vec{y} = \begin{bmatrix}
\vec{b}\\
0
\end{bmatrix}$
to obtain $y = \begin{bmatrix}
0\\
\vec{x}
\end{bmatrix}$.
%Therefore the rest of the article we assume that $M$ is Hermitian.\\
%The idea of the method is to reduce given MBVP to the system of linear algebraic equations \ref{mateq} and apply HHL algorithm.
%In this study we will consider an ideal case and refer reader to \cite{Dervovic2018QuantumLS} and literature therein for different methods of Hamiltonian simulation and quantum phase estimation.

As a result, we'll assume M is Hermitian for the rest of the article.
We will assume an ideal case in this study and refer the reader to \cite{Dervovic2018QuantumLS} and related literature for other approaches of Hamiltonian simulation and quantum phase estimation.\\
Stefan type Free Boundary Value Problems (FBVPs) \citep{crank1984free, gupta2017the, goldman1997inverse,figueiredo2007free} that take phase transformations into account can be used for modeling processes stated in \cite{friedman1970,rubinstein2000stefan,kharin2012} and agree with experimental data \cite{Khar, Sar_GH_P_0}. These problems are among the most difficult in the theory of non-linear parabolic equations where along with the temperature function they require determination of an unknown moving boundary function (Direct Stefan Problem) or a flux function (Inverse Stefan Problem). In some specific cases Heat Potentials can be constructed which allow boundary value problems to be reduced to integral equations \cite{friedman1970,rubinstein2000stefan,tikhonov2013equations}.
However, there are additional challenges in the case of domains that degenerate at the initial time due to the singularity of integral equations, which belong to the family of pseudo-Volterra equations and which are difficult to solve in the general case. The reader can refer to the extensive list of papers and literature on the FBVPs in \cite{Tarzia2000ABO}. Despite the importance and comprehensiveness of all of these studies, developing exact and approximate methods for solving FBVPs capable adequately model electric contact phenomena still remains a challenge in the theory of partial differential equations and mathematical physics.

%However, in the case of domains that degenerate at the initial time, there are additional difficulties because of the singularity of integral equations, which belong to the class of pseudo - Volterra equations which are unsolvable in the general case. A reader can refer to the long list of studies in \cite{Tarzia2000ABO} and literature therein dedicated to the MBVPs. Despite the great value and exhaustiveness of all these results, investigation and elaboration of both exact and approximate methods for solving MBVPs responsible for adequate modeling electric contact phenomena  is still an actual mathematical problem.

In this paper we consider a class of PDEs with free boundaries
\begin{align}
&\frac{\partial {\theta}}{\partial {t}} = a^2\left(\frac{\partial {\theta^2}}{\partial {x^2}} + \frac{\nu}{x}\frac{\partial{\theta}}{\partial{x}} \right),\hspace{1em} \alpha(t)<x<\beta(t),\hspace{1em}  -\infty<\nu<\infty,\hspace{1em} t>0, \label{geneq}\
\end{align}
which can be solved by the series of linear combinations of special functions which a priori satisfy the equation \ref{geneq}
\begin{align}
&S_{\gamma,\nu}^{1}\left(x,t\right) = \left(2a\sqrt{t}\right)^{\gamma}\Phi\left(-\frac{\gamma}{2},\frac{\nu+1}{2};-\frac{x^{2}}{4a^{2}t}\right),\hspace{1em} -\infty<\gamma,\nu<\infty,\hspace{1em} \label{hyper1}\\[1mm]
&S_{\gamma,\nu}^{2}\left(x,t\right) = \left(2a\sqrt{t}\right)^{\gamma}\left(\frac{x^{2}}{4a^{2}t}\right)^{\frac{1-\nu}{2}}\Phi\left(\frac{1-\nu-\gamma}{2},\frac{3-\nu}{2};-\frac{x^{2}}{4a^{2}t}\right),\label{hyper2}\hspace{1em}\
\end{align}
and gain exponential speed up when quantum HHL algorithm applied.
%Pioneering studies \cite{ben80,fe82} in 1980s gave a birth to a new paradigm in computation which we call nowadays quantum computing, whereby information is encoded in a quantum system. Further on, in 1990s a series of studies \cite{sho97,gro96,gro97} dedicated to quantum algorithms provided exponential speed-up in run time over the best known classical algorithms for same tasks. In last decades, consistent advances in theory and experiments generated a plethora of powerful quantum algorithms \cite{mo16} which surpass their classical counterparts in terms of computational power, however worth noting that their applications are restricted to few use cases due to the challenges related to their physical realization. Careful physical realization may lead to profound results in reaching exponential speed-up.\\

\begin{algorithm}[H]
\SetAlgoLined
\KwData{Load the data $\ket{b}\in\mathbb{C} ^{N}$}
\KwResult{Apply an observable  M  to calculate $F(x)=\bra{x}M\ket{x}.$}
initialization\;
\While{outcome is not $1$ }{
\begin{itemize} 
  \item Apply Quantum Phase Estimation (QPE) with\\ $U=e^{iMt}:=\sum_{j=0}^{N-1}e^{i\lambda_{j}t}\ket{u_j}\bra{u_j}$. Which implies $\sum_{j=0}^{N-1}b_{j}\ket{\lambda_j}_{n_{l}}\bra{u_j}_{n_{b}}$, \\ in the eigenbasis of  $M$\\
where $\ket{\lambda_j}_{n_{l}}$ is the  $n_{l}$ -bit binary representation of $\lambda_{j}$ .
  \item Add an ancilla qubit and apply a rotation conditioned on  $\ket{\lambda_j},$\\
  $\sum_{j=0}^{N-1}b_{j}\ket{\lambda_j}_{n_{l}}\bra{u_j}_{n_{b}}\left(\sqrt{1-\frac{C^2}{\lambda_j^2}}\ket{0}+\frac{C}{\lambda_j}\ket{1}\right)$, $C$ - normalization constant.
  \item Apply $QPE^{\dag}.$ This results in\\
  $\sum_{j=0}^{N-1}b_{j}\ket{0}_{n_{l}}\bra{u_j}_{n_{b}}\left(\sqrt{1-\frac{C^2}{\lambda_j^2}}\ket{0}+\frac{C}{\lambda_j}\ket{1}\right);$
\end{itemize}
\eIf{If the outcome is  $1$ , the register is in the post-measurement state\\
    $\left(\sqrt{\frac{1}{\sum_{j=1}^{N-1}\left|b_j\right|^2/\left|\lambda_j\right|^2/}}\right)\sum_{j=0}^{N-1}\frac{b_{j}}{\lambda_{j}}\ket{0}_{n_{l}}\bra{u_j}_{n_{b}}$ }{Apply an observable  M  to calculate $F(x)=\bra{x}M\ket{x}$;
}{
repeat the loop\;
}
}
\caption{Quantum HHL Algorithm in Qiskit}
\label{alg:genhhl}
\end{algorithm}
\begin{figure}[htbp]

\begin{quantikz}[row sep={1cm,between origins}, slice
titles=slice \col,slice style=blue,slice label style
={inner sep=1pt,anchor=south west,rotate=40}]
\lstick{$\ket{0}$}& \qw & \qw & \qw\slice{step 1} & \qw\slice{step 2}\gategroup[4,steps=3,style={dashed,
rounded corners,fill=blue!20, inner xsep=2pt},
background, label style={label position=below,anchor=
north,yshift=-0.2cm}]{{repeat until success}} & \gate[3]{\begin{turn}{90} 
Eigenvalue inversion
\end{turn}}\slice{step 3} & \qw\slice{step 4} & \meter{}\slice{step 5} & \qw & \qw & \qw \\
\lstick{$\ket{0}$}& \qwbundle
{} & \qw & \qw & \gate[3]{\begin{turn}{90} 
$QPE$
\end{turn}} & \qw &  \gate[3]{\begin{turn}{90} 
$QPE\dagger$
\end{turn}}& \qw & \qw & \qw & \qw \\
\lstick{$\ket{0}$}& \qwbundle
{} & \gate[2]{\begin{turn}{90} 
Load \ket{b}
\end{turn}} & \qw & \qw & \qw & \qw & \qw & \gate[2]{\begin{turn}{90} 
$F(x)$
\end{turn}} & \qw & \qw \\
\lstick{$\ket{0}$}& \qwbundle
{} & \qw & \qw & \qw & \qw & \qw & \qw & \qw &  \meter{}\slice{step 6} & \qw 
\end{quantikz}
\caption{Quantum HHL algorithm.}
\label{fig}
\end{figure}
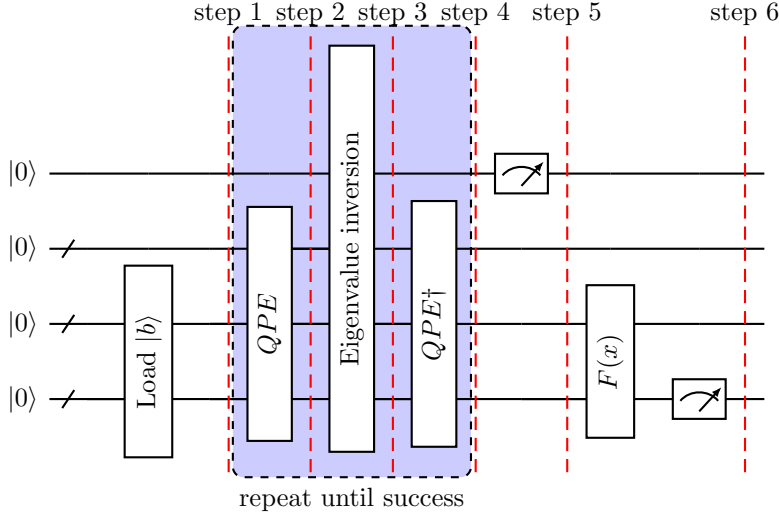

\section{Main results}
\label{sec:main}
Equation \Ref{geneq} with arbitrary $\nu$ is a generalized heat equation which can serve as a model for bridging processes in electrical contacts with variable cross section. For special cases $\nu=0,1,2$ equation \ref{geneq} is transformed to the heat equation in linear, spherical and cylindrical coordinates respectively.\\

Following formulas will ease our further calculations

\begin{align}
&\underset{x\to 0}{\lim}\frac{1}{z^{\beta}}\Phi\left(-\frac{\beta}{2},\mu;-z^{2}\right) = \frac{\Gamma(\mu)}{\Gamma(\mu+\frac{\beta}{2})},
\label{lhopital}\\
&\frac{\partial\theta}{\partial x}=\sum_{n=0}^\infty\left(4a_1^2t\right)^n\Bigg[ \Bigg.A_n\left(\frac{-x}{4a_1^2t}\right) L_{n-1}^{\mu}\left(\frac{-x^2}{4a_{1}^2t}\right)+ B_n\left(\frac{x^2}{4a_1^2t}\right)^{-\mu}\left(\frac{2x}{4a_1^2t}\right)\Bigg( \Bigg.(1-\mu)\nonumber\\
&\Phi\left(1-\mu-n, 2-\mu,\frac{-x^2}{4a_{1}^2t}\right)-\left(\frac{-x^2}{4a_{1}^2t}\right)^{-\mu}\left(\frac{x^2}{4a_1^2t}\right)\Phi\left(2-\mu-n, 3-\mu,\frac{-x^2}{4a_{1}^2t}\right)\Bigg) \Bigg.\Bigg] \Bigg.,
\label{der_wrt_x}\\
&\left.\frac{\partial\theta}{\partial x}\right|_{x=\sqrt{t}}=\sum_{n=0}^\infty\left(4a_1^2\right)^n\left(2a_1\sqrt{t}\right)^{2n-1}\left(\frac{-\alpha}{a_1}\right)\Bigg[ \Bigg.A_n\left(L_{n-1}^{\mu}\left(\frac{-\alpha}{2a_1}\right)\right)+ B_n\Bigg( \Bigg.(\mu-1)\nonumber\\
&\left(\frac{\alpha^2}{4a_1^2}\right)^{-\mu}\Phi\left(1-\mu-n, 2-\mu,\frac{-
\alpha^2}{4a_{1}^2}\right)+(\mu-1)\left(\frac{\alpha^2}{4a_1^2}\right)^{-\mu}\Phi\left(1-\mu-n, 2-\mu,\frac{-\alpha^2}{4a_{1}^2}\right)\Bigg) \Bigg.\Bigg] \Bigg.,
\label{der_of_theta}
\end{align}
\subsection{HHL algorithm for exact solution of one phase Inverse Stefan Problem with variable cross section domain}
\label{sec:onephase_isp}

Let's consider following one phase inverse Stefan Problem with variable cross section which is used for modeling arcing and bridging processes in electrical contacts. 
\begin{align}
&\frac{\partial {\theta}}{\partial {t}} = a^2\left(\frac{\partial {\theta^2}}{\partial {x^2}} + \frac{\nu}{x}\frac{\partial{\theta}}{\partial{x}} \right),\hspace{1em}  &0<x<\alpha(t),\hspace{1em} 0<\nu, t<1, \label{stef1_rob}\\[1mm]
&\theta(0,0) = T_{m},&\hspace{1mm}\alpha(0)=0,\label{stef1_incond1}\\[3mm]
&\left. \left(\beta \theta+\gamma \frac{\partial \theta}{\partial x} \right)\right|_{x=0 }= P(t),\label{stef1_flux}\\[3mm]
&\theta(\alpha(t),t) = T_m,\label{stef1_temp}\\[3mm]
&\left.\lambda\frac{\partial \theta}{\partial x} \right|_{x=\alpha \sqrt{t}} = L\gamma\frac{\partial \alpha(t)}{\partial t}\label{stef1_stefcond}
\end{align}
the solution can be represented in the following form
\begin{align}
&\theta(x,t)= \sum_{n=0}^{\infty}\left(4a^2t\right)^{n}\left[A_nL_{n}^{\mu-1}\left( \frac{-x^2}{4a^2t} \right)+ B_n\left( \frac{x^2}{4a^2t} \right)^{1-\mu}\Phi\left(1-\mu-n, 2-\mu,\frac{-x^2}{4a^2t}\right)\right] \label{solstef1}
\end{align}
where $k$ is defined from boundary conditions.

From conditions \ref{stef1_flux}, \ref{stef1_temp} and \ref{stef1_stefcond} we get following expressions
\begin{subequations}
\begin{align}
&\beta\sum_{n=0}^\infty\left(4a_1^2t\right)^nA_nL_n^{\mu-1}(0)+\sum_{n=0}^\infty P^n(0)\frac{t^n}{n!}=0,
\label{stef1_flux_sys}\\
&\sum_{n=0}^{\infty}\left(4a_{1}^2t\right)^{n}\left[A_nL_{n}^{\mu-1}\left( \frac{-\alpha^2}{4a_{1}^2} \right)+ B_n\left( \frac{\alpha^2}{4a_{1}^2} \right)^{1-\mu}\Phi\left(1-\mu-n, 2-\mu,\frac{-\alpha^2}{4a_{1}^2}\right)\right]=T_m,
\label{stef1_temp_sys}\\
&\sum_{n=0}^\infty\left(4a_1\right)^{n}\left(\sqrt{t}\right)^{2n-1}\left(\frac{-\alpha}{a_1}\right)\Bigg[ \Bigg.A_n\left(L_{n-1}^{\mu}\left(\frac{-\alpha}{2a_1}\right)\right)\nonumber + B_n\Bigg( \Bigg.(\mu-1)\left(\frac{\alpha^2}{4a_1^2}\right)^{-\mu}\Phi\left(1-\mu-n, 2-\mu,\frac{-
\alpha^2}{4a_{1}^2}\right)\nonumber\\
&+(\mu-1)\left(\frac{\alpha^2}{4a_1^2}\right)^{-\mu}\Phi\left(1-\mu-n, 2-\mu,\frac{-\alpha^2}{4a_{1}^2}\right)\Bigg) \Bigg.\Bigg] \Bigg.=L\gamma\frac{1}{2\sqrt{t}}
\label{stef1_stefcond_sys}
\end{align}
\end{subequations}
After multiplying both sides of \ref{stef1_stefcond_sys} by $\sqrt{t}$ and comparing coefficients at same powers of $t$ in \ref{stef1_flux_sys},\ref{stef1_temp_sys} and \ref{stef1_stefcond_sys} problem \ref{stef1_rob}-\ref{solstef1}  is reduced to the equation\ref{mateq}.
%\begin{align}
%&M=\begin{pmatrix}m_{11}&0&m_{13}&0&\dots&0\\
%m_{21}&m_{22}&0&0&\dots&0\\
%m_{31}&m_{32}&0&0&\dots&0\\
%\vdots&&&\ddots&&\vdots&\\
%0&\dots&0&m_{3k+13k+1}&0&m_{3k+13k+3}\\
%0&\dots&0&m_{3k+23k+1}&m_{3k+23k+2}&0\\
%0&\dots&0&m_{3(k+1)3k+1}&m_{3(k+1)3k+2}&0\end{pmatrix},\label{mentries}
%\\&x=\begin{pmatrix}A_0\\
%B_0\\
%P_0\\
%\vdots\\
%A_k\\
%B_k\\
%P_k\end{pmatrix}, 
%b=\begin{pmatrix} 0\\
%T_m\\\frac{L\gamma}{2}\\
%\vdots\\
%0\\
%0\\
%0
%\end{pmatrix}\nonumber,
%\end{align}
%where
%\begin{align*}
%&m_{3k+1,3k+1}=\beta\left(4a_1^2\right)^n L_n^{\mu -1}\left(0\right), 
%\\&m_{3k+1,3k+3}=\frac{P_n(0)}{n!},
%m_{3k+2,3k+1}= \left(4a_1^2\right)^n L_n^{\mu -1}\left(\frac{-\alpha^2}{4a_1^2}\right),\\
%&m_{3k+2,3k+2}= \left(4a_1^2\right)^n\left( \frac{\alpha^2}{4a_1^2} \right)^{1-\mu}\Phi\left(1-\mu-n, 2-\mu,\frac{-\alpha^2}{4a_1^2}\right),\\
%&m_{3k+3,3k+1}=\left(4a_1^2\right)^n \left(\frac{-\alpha}{a_1}\right) L_{n+1}^{\mu}\left(\frac{-\alpha^2}{4a_1^2}\right),\\
%&m_{3k+3,3k+2}= (\mu-1)\left(\frac{\alpha^2}{4a_1^2}\right)^{-\mu}\Phi\left(1-\mu-n, 2-\mu,\frac{-
%\alpha^2}{4a_{1}^2}\right)\nonumber\\
%&+(\mu-1)\left(\frac{\alpha^2}{4a_1^2}\right)^{1-\mu}\Phi\left(2-\mu-n, 3-\mu,\frac{-\alpha^2}{4a_{1}^2}\right)
%\end{align*}
%where entries $m_{i,j}$ are numbers which can be determined from tables or by calculators.
%\\
\subsubsection{Application for the calculation of the arc heat flux for melting of
a micro-asperity at electrical contact closure in vacuum circuit breakers}

\begin{figure}[h]
\centerline{\includegraphics[width=5cm, height=5cm]{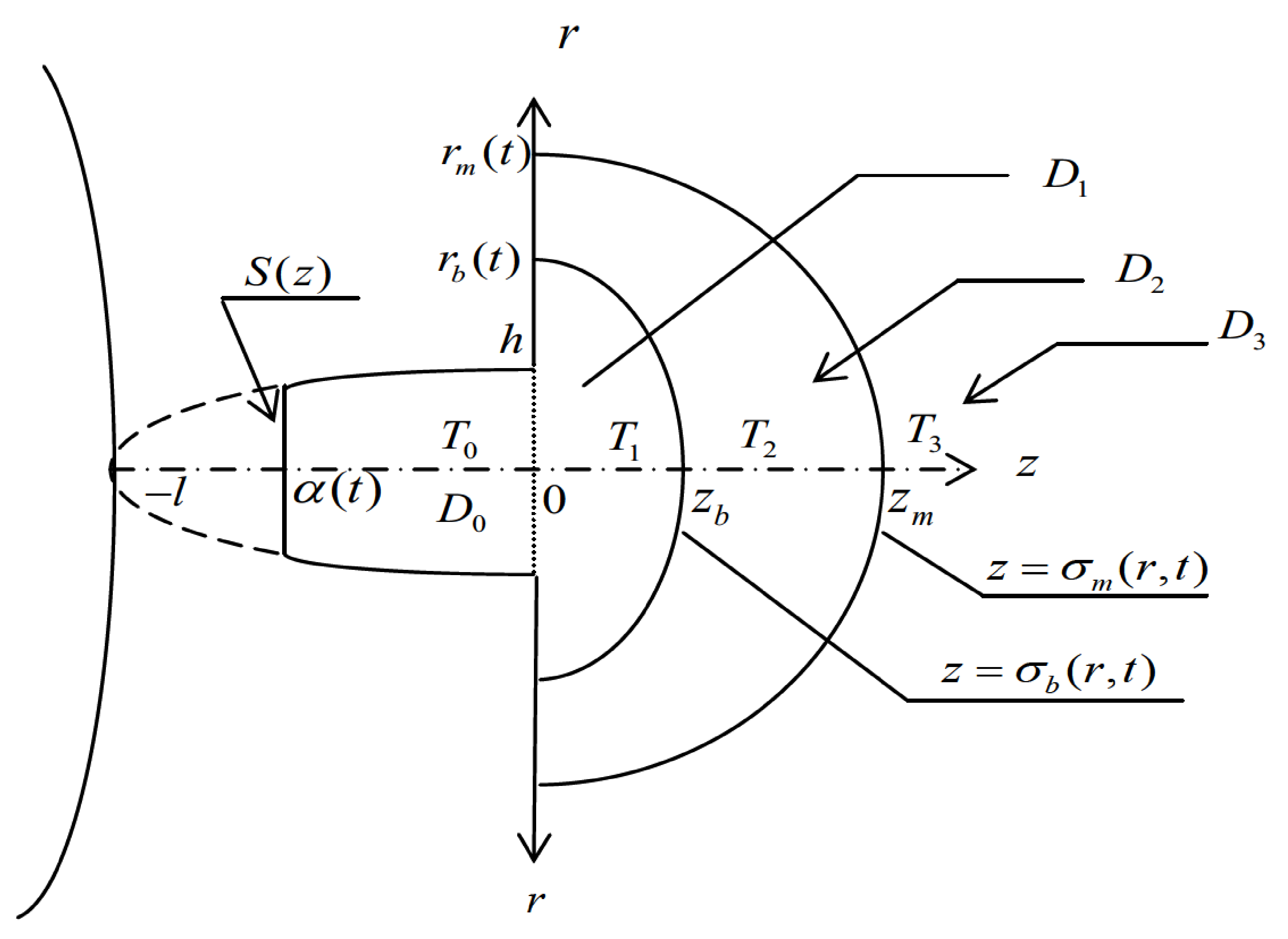}}
\caption{Micro-asperity}
\label{Microasperity}
\end{figure}

On the Figure \ref{Microasperity}, problem \ref{geneq_1}-\ref{concordance_1} models temperature and contact dynamics in region $D_0$ at the initial stage before mechanical bounce where $D_{0}[-l \leq z \leq 0]$ and $D_{1}\left[0 \leq z \leq \sigma_{b}(r, t)\right]$ are the zones of temperature dynamics and pressure of metal vapours in contact gap, including the zone of evaporated micro-asperity  with adjoining evaporated area inside contact, $D_{2}\left[\sigma_{b}(r, t) \leq z \leq \sigma_{m}(r, t)\right]$ melted zone and $D_{3}\left[\sigma_{m}(r, t) \leq z<\infty\right]$  is the solid zone. \\
The initial stage of heating and melting of a micro-asperity up to boiling temperature can be described by the bridge model, which is presented and discussed in details in \cite{1422606}. This liquid bridge can be considered as a bar with variable cross-section $S(z)$.
The shape of the cross-section $S(z)$ is chosen from the analysis of a Talyrond trace (profilogram) of the contact surface and in this case it's identified with a paraboloid having the altitude $l$ and the radius of the base $h$.
\\
Let us consider a model problem which is responsible for modeling initial stage of heating and melting of micro asperity in region $D_0$. 

\begin{align}
\frac{\partial \theta}{\partial t}=a^{2}\left(\frac{\partial^{2} \theta}{\partial r^{2}}+\frac{v}{r} \frac{\partial \theta}{\partial r}\right),
\label{geneq_1}
\end{align}

the domain
$$
D_0: \quad 0 \leq r<\alpha(t), \quad 0<t<T,
$$
boundary condition at $r=\alpha(t)$ 
\begin{align}
\theta(\alpha(t), t)=f(t),
\label{geneq_1_bc}
\end{align}

and the Stefan condition
\begin{align}
-\lambda \frac{\partial \theta(\alpha(t), t)}{\partial r}=P(t)-L \gamma \frac{d \alpha}{d t}.
\label{geneq_1_flux}
\end{align}
At the initial time the domain $D_0$ collapses into zero, thus the initial and concordance conditions are
\begin{align}
\alpha(0)=\theta(0,0)=f(0)=0.
\label{concordance_1}
\end{align}
It is required to find the functions $\theta(r, t)$ and $P(t)$, where $\alpha(t)$ and $f(t)$ are given.

%The process of Joule heating and evaporation of a micro-asperity and adjoining zone of contact material should be considered in dynamics step by step beginning from the time of contact touch. 

We represent the Heat Temperature function of problem \ref{geneq_1}-\ref{concordance_1} in the form
\begin{align}
    \theta(r, t)=\sum_{n=0}^{\infty} A_{n}\left(4 a^{2} t\right)^{n} L_{n}^{(\beta)}\left(-\frac{r^{2}}{4 a^{2} t}\right)
\label{sol_geneq_1}    
\end{align}
where $\beta=\frac{v-1}{2}$ and can be found like it's demonstrated in \cite{1422606}, $L_{n}(x)$ are associated Laguerre polynomials.
The function \Ref{sol_geneq_1} satisfies the heat equation \ref{geneq_1} for arbitrary constants $A_{n}$.
Satisfying the boundary condition \ref{geneq_1_bc} we get \begin{align}
f(t)=\sum_{n=0}^{\infty} A_{n}\left(4 a^{2}\right)^{n} L_{n}^{\beta}\left(-\frac{\alpha^{2}}{4 a^{2}}\right)t^{n},
\end{align}
\begin{align}
\sum_{n=0}^{\infty} \frac{f^{(n)}(0)}{n!} t^{n}=\sum_{n=0}^{\infty} A_{n}\left(4 a^{2}\right)^{n} L_{n}\left(-\frac{\alpha^{2}}{4 a^{2}}\right) t^{n},
\label{geneq_1_bc1}
\end{align}
%\begin{align}
 %A_{n}=\frac{f^{(n)}(0)}{n !\left(4 a^{2}\right)^{n} L_{n}^{\beta}\left(-\frac{\alpha^{2}}{4 a^{2}}\right)},\\
%L_{n}^{\beta}\left(-\frac{\alpha^{2}}{4 a^{2}}\right) \neq 0,\end{align}
where
\begin{align}
L^{\beta}_n\left(-\frac{\alpha^{2}}{4 a^{2}}\right)=\frac{\left(\beta+1\right)_{n}}{n!} \Hypergeometric{1}{1}{-n}{\beta+1}{t}
\nonumber
\end{align}

\begin{align}
    \left(\beta\right)_{n}=\frac{\Gamma(\beta+n)}{\Gamma(\beta)}=\frac{\int_{0}^{\infty}x^{\beta-n-1} e^{-x} d x}{\int_{0}^{\infty} x^{\beta-1} e^{-x} d x}
\nonumber
\end{align}
where $(\beta)_{n}-\text { Pochhammer  symbol}$
Equation \Ref{geneq_1_bc1} can be reduced to the equation \Ref{mateq} $A_n$ coefficients of Heat temperature function \Ref{geneq_1} can be determined.
The unknown heat flux $P(t)$ can be found now from the expression \ref{geneq_1_flux} which can be written in the form
\begin{align}
P(t)=\lambda \frac{\partial \theta (\alpha(t), t)}{\partial r}-L \gamma \frac{d \alpha}{d t}
\end{align}

where the right side is already found.
\\
\subsection{HHL algorithm for exact solution of two phase Inverse Stefan Problems with variable cross section domain}
Let's consider an inverse two phase Stefan problem. An inverse two phase Stefan problem with variable cross section is used for modeling bridging and arcing phenomena with phase transition predominantly from solid to liquid or vice-versa.
\label{sec:twophase_isp}
\begin{align}
&\frac{\partial {\theta_{1}}}{\partial {t}} = a_{1}^2\left(\frac{\partial {\theta_{1}^2}}{\partial {x^2}} + \frac{\nu}{x}\frac{\partial{\theta_{1}}}{\partial{x}} \right),\hspace{1em}  &0<x<\beta(t),\hspace{1em} 0<\nu, t<1 \label{stef2_eq_1}\\[1mm]
&\frac{\partial {\theta_{2}}}{\partial {t}} = a_{2}^2\left(\frac{\partial {\theta_{2}^2}}{\partial {x^2}} + \frac{\nu}{x}\frac{\partial{\theta_{2}}}{\partial{x}} \right),\hspace{1em}  &\beta(t)<x<\infty,\hspace{1em} \label{stef2_eq_2}\\[3mm]
&\theta_{1}(0,0) = T_{m},\label{stef2_incond1}\\[3mm]
&\theta_{2}(x,0) = f(x),\label{stef2_incond2}\\[3mm]
&f(0)=T_{m},\hspace{1mm}\alpha(0)=0,\hspace{1mm}\underset{x\to \infty}{\lim}f(x)\approx f(X)=0,\hspace{1mm} \hspace{1mm}&\underset{x\to \infty}{\lim}\theta(x,t)\approx \theta(X,t)=0,\label{stef2_concord}\\[3mm]
&\left. \left(\beta \theta+\gamma \frac{\partial \theta}{\partial x} \right)\right|_{x=0 }= P(t),
\label{stef2_flux}\\[3mm]
&\theta_{1}(\alpha(t),t) = \theta_{2}(\alpha(t),t)=T_{m}  \label{stef2_temp}\\[3mm]
&\left.-\lambda_{1}\frac{\partial \theta_{1}}{\partial t} \right|_{x=\alpha(t)} = \left.-\lambda_{2}\frac{\partial \theta_{2}}{\partial x} \right|_{x=\alpha(t)}+L\gamma\frac{\partial \alpha(t)}{\partial x}\label{stef2_stefcond}
\end{align}
The solution of above problem \ref{stef2_eq_1}-\ref{stef2_stefcond} can be represented in the following form
\begin{align}
&\theta_{1}(x,t)= \sum_{n=0}^{\infty}\left(4a_{1}^2t\right)^{n}\left[A_nL_{n}^{\mu-1}\left( \frac{-x^2}{4a_{1}^2t} \right)+ B_n\left(4a_{1}^2t\right)^{1-\mu}\left( \frac{x^2}{4a_{1}^2t} \right)\Phi\left(1-\mu-n, 2-\mu,\frac{-x^2}{4a_{1}^2t}\right)\right] \label{sol1stef1}\\[1mm]
&\theta_{2}(x,t)= \sum_{n=0}^{\infty}\left(4a_{2}^2t\right)^{n}\left[C_nL_{n}^{\mu-1}\left( \frac{-x^2}{4a_{2}^2t} \right)+ D_n\left(4a_{2}^2t\right)^{1-\mu}\left( \frac{x^2}{4a_{2}^2t} \right)\Phi\left(1-\mu-n, 2-\mu,\frac{-x^2}{4a_{2}^2t}\right)\right] \label{sol2stef1}
\end{align}
From conditions \ref{stef1_flux}, \ref{stef1_temp} and \ref{stef1_stefcond} we get following expressions
\begin{subequations}
\begin{align}
&\frac{(-1)^n}{n!}C_n + D_n=\frac{f^{2n}(0)}{(2n)!},
\label{stef2_incond_C_D_n}\\
&\beta\sum_{n=0}^\infty\left(4a_1^2t\right)^nA_nL_n^{\mu-1}(0)+\sum_{n=0}^\infty P^n(0)\frac{t^n}{n!}=0,
\label{stef2_flux_sys}\\
&\sum_{n=0}^{\infty}\left(4a_{1}^2t\right)^{n}\left[A_nL_{n}^{\mu-1}\left( \frac{-\alpha^2(t)}{4a_{1}^2} \right)+ B_n\left( \frac{\alpha^2(t)}{4a_{1}^2} \right)^{1-\mu}\Phi\left(1-\mu-n, 2-\mu,\frac{-\alpha^2(t)}{4a_{1}^2}\right)\right]=T_m,
\label{stef2_temp1_sys}\\
&\sum_{n=0}^{\infty}\left(4a_{2}^2t\right)^{n}\left[C_nL_{n}^{\mu-1}\left( \frac{-\alpha^2(t)}{4a_{2}^2} \right)+ D_n\left( \frac{\alpha^2(t)}{4a_{2}^2} \right)^{1-\mu}\Phi\left(1-\mu-n, 2-\mu,\frac{-\alpha^2(t)}{4a_{2}^2}\right)\right]=T_m,
\label{stef2_temp2_sys}\\
&-\lambda_1\sum_{n=0}^\infty\left(4a_1^2t\right)^n\Bigg[ \Bigg.A_n\left(\frac{2\alpha(t)}{4a_1^2t}\right) L_{n-1}^{\mu}\left(\frac{-\alpha^2(t)}{4a_{1}^2t}\right) + B_n\left(\frac{\alpha^2(t)}{4a_1^2t}\right)^{-\mu}\left(\frac{2\alpha(t)}{4a_1^2t}\right)\Bigg( \Bigg.(1-\mu)\nonumber\\
&\Phi\left(1-\mu-n, 2-\mu,\frac{-\alpha^2(t)}{4a_{1}^2t}\right)-\left(\frac{\alpha^2(t)}{4a_{1}^2t}\right)\Phi\left(2-\mu-n, 3-\mu,\frac{\alpha^2(t)}{4a_{1}^2t}\right)\Bigg) \Bigg.\Bigg] \Bigg.\nonumber\\
+&\lambda_2\sum_{n=0}^\infty\left(4a_2^2t\right)^n\Bigg[ \Bigg.C_n\left(\frac{2\alpha(t)}{4a_2^2t}\right) L_{n-1}^{\mu}\left(\frac{-\alpha^2(t)}{4a_{2}^2t}\right)+ D_n\left(\frac{\alpha^2(t)}{4a_2^2t}\right)^{-\mu}\left(\frac{2\alpha(t)}{4a_2^2t}\right)\Bigg( \Bigg.(1-\mu)\nonumber\\
&\Phi\left(1-\mu-n, 2-\mu,\frac{\alpha^2(t)}{4a_{2}^2t}\right)-\left(\frac{-\alpha^2(t)}{4a_{2}^2t}\right)\Phi\left(2-\mu-n, 3-\mu,\frac{\alpha^2(t)}{4a_{2}^2t}\right)\Bigg) \Bigg.\Bigg] \Bigg.\nonumber\\
&=L\gamma\frac{d\alpha(t)}{dt}
\label{stef2_stefcond_sys}
\end{align}
\end{subequations}

The idea of the collocation method applied in this problem is to subdivide $0<t<T_a$ into $k$ intervals and after substituting solution functions \ref{sol1stef1}, \ref{sol2stef1} into the boundary conditions \ref{stef2_flux},\ref{stef2_temp},\ref{stef2_stefcond} at points $t_1, t_2,...,t_k$ reduce the problem to the problem \ref{mateq} where we apply HHL algorithm to determine coefficients $A_n, B_n, C_n, P_n$. For computational purposes we use Qiskit and IBM's quantum device.

\section{Experimental Results, Discussion and Conclusion}

%\subsection{Experimental Results on IBM Qiskit} 
%of \texorpdfstring{{\boldmath$Z=X \cup Y$}}{Z = X union Y}}
We used IBM Q \cite{IBMQ} and Qiskit for experiments and programming purposes. Inverse One and Two Phase Stefan Problems were solved demonstrating fidelity of 0.94. We refer reader to \cite{Sar_GH_P2} for details of experiments. 
Proposed method in combination with Fa Di Bruno's Formula and Quantum HHL algorithm can be used for exact solutions for direct/inverse Stefan type problems and MBVPs in general for arbitrary $\nu$ in \ref{geneq} and arbitrary $\alpha(t)$.
Special functions method in combination with HHL algorithm can be also used for approximate solutions of boundary value problems with fixed boundaries as well. Variational Quantum Linear Solver \citep{Bravo-Prieto} can be applied alternatively even though it's weaker in terms of computational power than HHL algorithm, it can be implemented on NISQ computers whereas HHL requires more robust computers.

%\section{Conclusion}

%\section*{References}

\bibliography{mybibfile}

\begin{thebibliography}{10}
\expandafter\ifx\csname url\endcsname\relax
  \def\url#1{\texttt{#1}}\fi
\expandafter\ifx\csname urlprefix\endcsname\relax\def\urlprefix{URL }\fi
\expandafter\ifx\csname href\endcsname\relax
  \def\href#1#2{#2} \def\path#1{#1}\fi

\bibitem{hhl09}
A.~W. Harrow, A.~Hassidim, S.~Lloyd,
  \href{https://link.aps.org/doi/10.1103/PhysRevLett.103.150502}{Quantum
  algorithm for linear systems of equations}, Phys. Rev. Lett. 103 (2009)
  150502.
\newblock \href {https://doi.org/10.1103/PhysRevLett.103.150502}
  {\path{doi:10.1103/PhysRevLett.103.150502}}.
\newline\urlprefix\url{https://link.aps.org/doi/10.1103/PhysRevLett.103.150502}

\bibitem{Haimo}
F.~M. Cholewinski, D.~T. Haimo,
  \href{http://www.jstor.org/stable/2027811}{Classical analysis and the
  generalized heat equation}, SIAM Review 10~(1) (1968) 67--80.
\newline\urlprefix\url{http://www.jstor.org/stable/2027811}

\bibitem{Haimo1}
D.~T. Haimo, \href{http://www.jstor.org/stable/2946288}{Series representation
  of generalized temperature functions}, SIAM Journal on Applied Mathematics
  15~(2) (1967) 359--367.
\newline\urlprefix\url{http://www.jstor.org/stable/2946288}

\bibitem{Widder}
D.~V. Widder, \href{http://www.jstor.org/stable/1998937}{The huygens property
  for the heat equation}, Transactions of the American Mathematical Society 232
  (1977) 239--244.
\newline\urlprefix\url{http://www.jstor.org/stable/1998937}

\bibitem{nasim_aggarwala_1984}
C.~Nasim, B.~D. Aggarwala, On the generalized heat-equation, Proceedings of the
  Edinburgh Mathematical Society 27~(3) (1984) 261–273.
\newblock \href {https://doi.org/10.1017/S0013091500022392}
  {\path{doi:10.1017/S0013091500022392}}.

\bibitem{ben80}
P.~{Benioff}, {The computer as a physical system: A microscopic quantum
  mechanical Hamiltonian model of computers as represented by Turing machines},
  Journal of Statistical Physics 22~(5) (1980) 563--591.
\newblock \href {https://doi.org/10.1007/BF01011339}
  {\path{doi:10.1007/BF01011339}}.

\bibitem{fe82}
R.~P. {Feynman}, {Simulating Physics with Computers}, International Journal of
  Theoretical Physics 21~(6-7) (1982) 467--488.
\newblock \href {https://doi.org/10.1007/BF02650179}
  {\path{doi:10.1007/BF02650179}}.

\bibitem{sho97}
P.~W. Shor, \href{https://doi.org/10.1137/S0097539795293172}{Polynomial-time
  algorithms for prime factorization and discrete logarithms on a quantum
  computer}, SIAM J. Comput. 26~(5) (1997) 1484–1509.
\newblock \href {https://doi.org/10.1137/S0097539795293172}
  {\path{doi:10.1137/S0097539795293172}}.
\newline\urlprefix\url{https://doi.org/10.1137/S0097539795293172}

\bibitem{gro96}
L.~K. Grover, \href{https://doi.org/10.1145/237814.237866}{A fast quantum
  mechanical algorithm for database search}, in: Proceedings of the
  Twenty-Eighth Annual ACM Symposium on Theory of Computing, STOC ’96,
  Association for Computing Machinery, New York, NY, USA, 1996, p. 212–219.
\newblock \href {https://doi.org/10.1145/237814.237866}
  {\path{doi:10.1145/237814.237866}}.
\newline\urlprefix\url{https://doi.org/10.1145/237814.237866}

\bibitem{gro97}
L.~K. Grover,
  \href{https://link.aps.org/doi/10.1103/PhysRevLett.79.325}{Quantum mechanics
  helps in searching for a needle in a haystack}, Phys. Rev. Lett. 79 (1997)
  325--328.
\newblock \href {https://doi.org/10.1103/PhysRevLett.79.325}
  {\path{doi:10.1103/PhysRevLett.79.325}}.
\newline\urlprefix\url{https://link.aps.org/doi/10.1103/PhysRevLett.79.325}

\bibitem{mo16}
A.~Montanaro, Quantum algorithms: an overview, npj Quantum Information 2
  (2016).

\bibitem{Preskill2018quantumcomputingin}
J.~Preskill, \href{https://doi.org/10.22331/q-2018-08-06-79}{Quantum
  {C}omputing in the {NISQ} era and beyond}, {Quantum} 2 (2018) 79.
\newblock \href {https://doi.org/10.22331/q-2018-08-06-79}
  {\path{doi:10.22331/q-2018-08-06-79}}.
\newline\urlprefix\url{https://doi.org/10.22331/q-2018-08-06-79}

\bibitem{Ritter_2019}
M.~B. Ritter, \href{https://doi.org/10.1088/1742-6596/1290/1/012003}{Near-term
  quantum algorithms for quantum many-body systems}, Journal of Physics:
  Conference Series 1290~(1) (2019) 012003.
\newblock \href {https://doi.org/10.1088/1742-6596/1290/1/012003}
  {\path{doi:10.1088/1742-6596/1290/1/012003}}.
\newline\urlprefix\url{https://doi.org/10.1088/1742-6596/1290/1/012003}

\bibitem{Supremacy}
H.-S. Zhong, H.~Wang, Y.-H. Deng, M.-C. Chen, L.-C. Peng, Y.-H. Luo, J.~Qin,
  D.~Wu, X.~Ding, Y.~Hu, P.~Hu, X.-Y. Yang, W.-J. Zhang, H.~Li, Y.~Li,
  X.~Jiang, L.~Gan, G.~Yang, L.~You, Z.~Wang, L.~Li, N.-L. Liu, C.-Y. Lu, J.-W.
  Pan, \href{https://www.science.org/doi/abs/10.1126/science.abe8770}{Quantum
  computational advantage using photons}, Science 370~(6523) (2020) 1460--1463.
\newblock \href
  {http://arxiv.org/abs/https://www.science.org/doi/pdf/10.1126/science.abe8770}
  {\path{arXiv:https://www.science.org/doi/pdf/10.1126/science.abe8770}}, \href
  {https://doi.org/10.1126/science.abe8770}
  {\path{doi:10.1126/science.abe8770}}.
\newline\urlprefix\url{https://www.science.org/doi/abs/10.1126/science.abe8770}

\bibitem{Harrow_2017}
A.~W. Harrow, A.~Montanaro,
  \href{https://doi.org/10.1038%2Fnature23458}{Quantum computational
  supremacy}, Nature 549~(7671) (2017) 203--209.
\newblock \href {https://doi.org/10.1038/nature23458}
  {\path{doi:10.1038/nature23458}}.
\newline\urlprefix\url{https://doi.org/10.1038%2Fnature23458}

\bibitem{Montanaro_2016}
A.~Montanaro, S.~Pallister,
  \href{https://doi.org/10.1103%2Fphysreva.93.032324}{Quantum algorithms and
  the finite element method}, Physical Review A 93~(3) (mar 2016).
\newblock \href {https://doi.org/10.1103/physreva.93.032324}
  {\path{doi:10.1103/physreva.93.032324}}.
\newline\urlprefix\url{https://doi.org/10.1103%2Fphysreva.93.032324}

\bibitem{Wiebe}
N.~{Wiebe}, D.~{Braun}, S.~{Lloyd}, {Quantum Algorithm for Data Fitting},
  Physical Review Letters 109~(5) (2012) 050505.
\newblock \href {http://arxiv.org/abs/1204.5242} {\path{arXiv:1204.5242}},
  \href {https://doi.org/10.1103/PhysRevLett.109.050505}
  {\path{doi:10.1103/PhysRevLett.109.050505}}.

\bibitem{Lloyd}
S.~Lloyd, M.~Mohseni, P.~Rebentrost,
  \href{https://arxiv.org/abs/1307.0411}{Quantum algorithms for supervised and
  unsupervised machine learning} (2013).
\newblock \href {https://doi.org/10.48550/ARXIV.1307.0411}
  {\path{doi:10.48550/ARXIV.1307.0411}}.
\newline\urlprefix\url{https://arxiv.org/abs/1307.0411}

\bibitem{Rebentrost}
P.~Rebentrost, M.~Mohseni, S.~Lloyd,
  \href{https://arxiv.org/abs/1307.0471}{Quantum support vector machine for big
  data classification} (2013).
\newblock \href {https://doi.org/10.48550/ARXIV.1307.0471}
  {\path{doi:10.48550/ARXIV.1307.0471}}.
\newline\urlprefix\url{https://arxiv.org/abs/1307.0471}

\bibitem{Barz2014-fo}
S.~Barz, I.~Kassal, M.~Ringbauer, Y.~O. Lipp, B.~Daki{\'c}, A.~Aspuru-Guzik,
  P.~Walther, A two-qubit photonic quantum processor and its application to
  solving systems of linear equations, Scientific Reports 4~(1) (2014) 6115.

\bibitem{Cai}
X.-D. Cai, C.~Weedbrook, Z.-E. Su, M.-C. Chen, M.~Gu, M.-J. Zhu, L.~Li, N.-L.
  Liu, C.-Y. Lu, J.-W. Pan,
  \href{https://link.aps.org/doi/10.1103/PhysRevLett.110.230501}{Experimental
  quantum computing to solve systems of linear equations}, Phys. Rev. Lett. 110
  (2013) 230501.
\newblock \href {https://doi.org/10.1103/PhysRevLett.110.230501}
  {\path{doi:10.1103/PhysRevLett.110.230501}}.
\newline\urlprefix\url{https://link.aps.org/doi/10.1103/PhysRevLett.110.230501}

\bibitem{Zheng}
Y.~Zheng, C.~Song, M.-C. Chen, B.~Xia, W.~Liu, Q.~Guo, L.~Zhang, D.~Xu,
  H.~Deng, K.~Huang, Y.~Wu, Z.~Yan, D.~Zheng, L.~Lu, J.-W. Pan, H.~Wang, C.-Y.
  Lu, X.~Zhu,
  \href{https://link.aps.org/doi/10.1103/PhysRevLett.118.210504}{Solving
  systems of linear equations with a superconducting quantum processor}, Phys.
  Rev. Lett. 118 (2017) 210504.
\newblock \href {https://doi.org/10.1103/PhysRevLett.118.210504}
  {\path{doi:10.1103/PhysRevLett.118.210504}}.
\newline\urlprefix\url{https://link.aps.org/doi/10.1103/PhysRevLett.118.210504}

\bibitem{Pan_PhysRevA}
J.~Pan, Y.~Cao, X.~Yao, Z.~Li, C.~Ju, H.~Chen, X.~Peng, S.~Kais, J.~Du,
  \href{https://link.aps.org/doi/10.1103/PhysRevA.89.022313}{Experimental
  realization of quantum algorithm for solving linear systems of equations},
  Phys. Rev. A 89 (2014) 022313.
\newblock \href {https://doi.org/10.1103/PhysRevA.89.022313}
  {\path{doi:10.1103/PhysRevA.89.022313}}.
\newline\urlprefix\url{https://link.aps.org/doi/10.1103/PhysRevA.89.022313}

\bibitem{Dervovic2018QuantumLS}
D.~Dervovic, M.~Herbster, P.~Mountney, S.~Severini, N.~Usher, L.~Wossnig,
  Quantum linear systems algorithms: a primer, ArXiv abs/1802.08227 (2018).

\bibitem{crank1984free}
J.~Crank, Free and moving boundary problems, Clarendon Press, Oxford
  Oxfordshire New York, 1984.

\bibitem{gupta2017the}
S.~C. Gupta, The Classical Stefan Problem, Elsevier Science, Amsterdam, 2017.

\bibitem{goldman1997inverse}
N.~L. Goldman, Inverse Stefan problems, Kluwer Academic Publishers, Dordrecht
  Boston, 1997.

\bibitem{figueiredo2007free}
I.~Figueiredo, Free boundary problems : theory and applications, Birkhäuser
  Verlag, Basel, Switzerland Boston Ma, 2007.

\bibitem{friedman1970}
A.~Friedman, \href{https://projecteuclid.org:443/euclid.bams/1183532190}{Free
  boundary problems for parabolic equations}, Bull. Amer. Math. Soc. 76~(5)
  (1970) 934--941.
\newline\urlprefix\url{https://projecteuclid.org:443/euclid.bams/1183532190}

\bibitem{rubinstein2000stefan}
L.~Rubinstein, \href{https://books.google.com/books?id=rmPLCwAAQBAJ}{The Stefan
  Problem}, Translations of Mathematical Monographs, American Mathematical
  Society, 2000.
\newline\urlprefix\url{https://books.google.com/books?id=rmPLCwAAQBAJ}

\bibitem{kharin2012}
S.~Kharin, M.~Sarsengeldin, Influence of contact materials on phenomena in a
  short electrical arc, in: Advanced Materials XII, Vol. 510 of Key Engineering
  Materials, Trans Tech Publications Ltd, 2012, pp. 321--329.
\newblock \href {https://doi.org/10.4028/www.scientific.net/KEM.510-511.321}
  {\path{doi:10.4028/www.scientific.net/KEM.510-511.321}}.

\bibitem{Khar}
S.~N. Kharin, Role of metallic vapor pressure in contact bouncing and welding
  at closure of electrical contacts in vacuum, in: 2012 IEEE 58th Holm
  Conference on Electrical Contacts (Holm), 2012, pp. 1--7.

\bibitem{Sar_GH_P_0}
{Merey Sarsengeldin},
  \href{https://github.com/users/Schrodinger-cat-kz/projects/2}{Special
  functions and hhl quantum algorithm for solving moving boundary value
  problems occurring in electric contact phenomena} (2020).
\newline\urlprefix\url{https://github.com/users/Schrodinger-cat-kz/projects/2}

\bibitem{tikhonov2013equations}
A.~Tikhonov, A.~Samarskii,
  \href{https://books.google.com/books?id=PTmoAAAAQBAJ}{Equations of
  Mathematical Physics}, Dover Books on Physics, Dover Publications, 2013.
\newline\urlprefix\url{https://books.google.com/books?id=PTmoAAAAQBAJ}

\bibitem{Tarzia2000ABO}
D.~A. Tarzia, A bibliography on moving-free boundary problems for the
  heat-diffusion equation. the stefan and related problems, Materials 2 (2000)
  1--297.

\bibitem{1422606}
S.~Kharin, H.~Nouri, D.~Amft, Dynamics of arc phenomena at closure of
  electrical contacts in vacuum circuit breakers, in: XXIst International
  Symposium on Discharges and Electrical Insulation in Vacuum, 2004.
  Proceedings. ISDEIV., Vol.~2, 2004, pp. 301--306.
\newblock \href {https://doi.org/10.1109/DEIV.2004.1422606}
  {\path{doi:10.1109/DEIV.2004.1422606}}.

\bibitem{IBMQ}
\href{https://quantum-computing.ibm.com/}{Ibm q 5 yorktown backend
  specification v1.1.0} (2018).
\newline\urlprefix\url{https://quantum-computing.ibm.com/}

\bibitem{Sar_GH_P2}
{Merey Sarsengeldin},
  \href{https://github.com/Schrodinger-cat-kz/P2_Hybrid-Classical-Quantum-Platform-for-solving-MBVPS.git}{A
  hybrid classical-quantum framework for solving free boundary value problems
  and applications in modeling electric contact phenomena.} (2020).
\newline\urlprefix\url{https://github.com/Schrodinger-cat-kz/P2_Hybrid-Classical-Quantum-Platform-for-solving-MBVPS.git}

\bibitem{Bravo-Prieto}
C.~Bravo-Prieto, R.~LaRose, M.~Cerezo, Y.~Subasi, L.~Cincio, P.~J. Coles,
  \href{https://arxiv.org/abs/1909.05820}{Variational quantum linear solver}
  (2019).
\newblock \href {https://doi.org/10.48550/ARXIV.1909.05820}
  {\path{doi:10.48550/ARXIV.1909.05820}}.
\newline\urlprefix\url{https://arxiv.org/abs/1909.05820}

\end{thebibliography}

\end{document}